# Engineering of SnO$_2$ – Graphene Oxide Nano-Heterojunctions for Selective Room-temperature Chemical Sensing and Optoelectronic Devices

Eleonora Pargoletti,[1,2] Umme H. Hossain,[3] Iolanda Di Bernardo,[4,†] Hongjun Chen,[4] Thanh Tran-Phu,[4] Gian Luca Chiarello,[1] Josh Lipton-Duffin,[5] Antonio Tricoli[4,*] and Giuseppe Cappelletti[1,2,*]

[1] Dipartimento di Chimica, Università degli Studi di Milano, via Golgi 19, 20133, Milano, Italy

[2] Consorzio Interuniversitario Nazionale per la Scienza e Tecnologia dei Materiali (INSTM), Via Giusti 9, 50121, Firenze, Italy

[3] Department of Electronic Materials Engineering, Research School of Physics and Engineering, The Australian National University, Canberra ACT 2601, Australia

[4] Nanotechnology Research Laboratory, College of Engineering and Computer Science, The Australian National University, Canberra ACT 2601, Australia

[5] Institute for Future Environments (IFE), Central Analytical Research Facility (CARF), Queensland University of Technology (QUT), Brisbane, Australia

*KEYWORDS Nano-heterojunctions; graphene oxide; tin dioxide; UV-photodetectors; room-temperature chemoresistive sensing; selectivity.*

**ABSTRACT:** The development of high-performing sensing materials, able to detect ppb-trace concentrations of volatile organic compounds at low temperatures, is required for the development of next-generation miniaturized wireless sensors. Here, we present the engineering of selective room-temperature chemical sensors, comprising of highly porous tin dioxide (SnO$_2$) – graphene oxide (GO) nano-heterojunction layouts. The optoelectronic and chemical properties of these highly porous (> 90%) p-n heterojunctions were systematically investigated in terms of composition and morphologies. Optimized SnO$_2$-GO layouts demonstrate significant potential as both visible-blind photodetectors and as selective room-temperature chemical sensors. Notably, a low GO content results in an excellent UV light responsivity (400 A·W$^{-1}$), with short rise and decay times, and room-temperature high chemical sensitivity with selective detection of volatile organic compounds such as ethanol down to 100 ppb. In contrast, a high concentration of GO drastically decreases the room-temperature response to ethanol, and results in good selectivity to ethylbenzene. The feasibility of tuning the chemical selectivity of the sensor response by engineering the relative amount of GO and SnO$_2$ is a promising feature that may guide the future development of miniaturized solid-state gas sensors. Furthermore, the excellent optoelectronic properties of these SnO$_2$-GO nano-heterojunctions may find applications to various other areas such as optoelectronic devices and (photo)electrocatalysis.

**Introduction**

The development of ultra-miniaturized and low-power consumption sensors for monitoring of volatile organic compounds (VOCs) concentrations is becoming increasingly important, due to the rapid pace of emission of potentially toxic VOCs in urban areas, and to their role as biomarkers in non-invasive medical diagnostics.[1–3] Many VOCs are highly toxic with potential carcinogenic, mutagenic and teratogenic function at low concentrations. They can also contribute to atmospheric pollution, such as photochemical smog and destruction of the ozone layer.[4] Recently, significant attention has been devoted to the BTEX compounds, namely benzene, toluene, ethylbenzene and xylol, due to their increasing release in various industrial processes.[4–6] Quite a few VOCs are also present in the human breath and correlated to several metabolic processes. Specifically, the monitoring of VOCs spontaneously released by the body is increasingly considered as promising path for non-invasive medical diagnostics and health monitoring.[7–9] For instance, abnormal concentrations of acetone (> 1800 ppb) can be related to type 1 diabetes, where standard concentrations in people not affected by this illness are 300-900 ppb.[10] Similarly, a marked presence of ethanol and acetone are related to non-alcoholic fatty liver disease and hepatic steatosis.[11] Ethylbenzene as well, apart from being a BTEX compound, has been recently recognized as one of the potential biomarkers for lung cancer detection (0.04 ppb in healthy humans *vs* 0.11 ppb in ill patients).[12–14] Hence, the need for frequent VOCs monitoring with deployable, portable or wearable detectors has attracted a widespread interest in the development of few-millimetres in size wireless sensor devices.[7]

Chemoresistive gas sensors, based on nanostructured metal oxides semiconductors (MOS), are a promising technology for low concentration detection of VOCs with superior miniaturization potential to established analytical techniques such as PTR-MS and gas chromatography.[15] Development of MOS sensors is held back from few fundamental challenges related to the sensing material, including high operating temperatures (200-400°C)[16–18] and their difficulty in achieving selectivity in multiple gas environments.[3] Much effort is focused to address the above challenges. Various recent studies report the design and fabrication of MOS (*e.g.* ZnO,[19–21] NiO,[16] WO$_3$[22]) with unique nanoarchitectures that enable low temperature sensing.[16–18] However, the lower limit of detection are often at the ppm level, and thus too high for many applications, including detection of important biomarkers for breath analysis.[17,18]

The use of heterojunctions between metal oxides[16,23] or by coupling them with carbonaceous materials[24,25] has been reported as a path to improve the gas sensing performance of MOS, with particular merits for room-temperature detection under light irradiation. Notably, graphene materials possess several promising features such as thermo-electric conductivity and mechanical strength,[12] which can enhance the sensing behavior of MOS by formation of nano-scale heterojunctions. Reduced graphene oxide (rGO) has been widely investigated and offers some potential for gas sensing.[26,27] In contrast, pure graphene oxide (GO) has been scarcely reported for this application,[28,29] due to its less defective structure and surface chemistry.[28,30,31] Nevertheless, its oxygen-rich functional groups can be the anchor points that help the further growth of MOS nanoparticles. Particularly, if the adopted metal oxide behaves as an n-type semiconductor, the conceivable formation of p(GO)[32–34]-n(MOS) heterojunction may be hypothesized.

Here, we report the fabrication of ultraporous nano-heterojunction networks of SnO$_2$ and GO, demonstrating the effective engineering of their chemical sensing and photoresponsive properties by the relative fraction of p- and n-type nanodomains. The photo- and chemical sensing properties of these nano-heterojunctions were systematically investigated, achieving insights into the role played by the graphene oxide in the enhancement of the room-temperature sensor response and the selectivity towards a particular VOC. Specifically, three different volatile compounds, *i.e.* ethanol (EtOH), acetone and ethylbenzene (EtBz), were adopted as target molecules. Notably, we observed that small amount of GO leads to the formation of electron depleted nano-heterojunctions with superior electron-hole separation efficiency. The latter nanostructures are able to selectively detect ethanol concentrations down to 100 ppb at room-temperature. In contrast, increasing the GO content hinders ethanol sensing and favors ethylbenzene detection, providing a mechanism to tailor MOS sensor selectivity. We demonstrated that this optimal nanocomposite structures provides excellent photo- and chemical responses, showcasing their applicability as both visible-blind UV photodetectors and selective room-temperature VOCs solid-state sensors.

**Experimental section**

All the chemicals were of reagent-grade purity and were used without further purification; doubly distilled water passed through a Milli-Q apparatus was utilized. All the reagents used were purchased from Sigma-Aldrich.

**Synthesis of pristine oxides and hybrid SnO$_2$-GO compounds**

Graphene Oxide (GO) was prepared by adopting a modified Hummers' method, already reported in the literature.[29,35,36] For the composite materials, *i.e.* SnO$_2$-GO, the adopted synthetic route was the same reported in our previous work[28] with different starting salt precursor-to-GO weight ratios (*i.e.* 4:1 and 32:1 SnO$_2$/GO, since with the other intermediate ratios lower sensing performances were obtained as deeply discussed in our previous study).[29] Specifically, the appropriate amount of SnCl$_4 \times$ 5H$_2$O was dissolved in 3.0 mg mL$^{-1}$ of an aqueous GO suspension. The mixture was stirred ($\omega$ = 300 rpm) for 3 h at 50 °C and, then, 30 mL of stoichiometric urea aqueous solution was added dropwise. The mixture was continuously stirred for other 3 h. Subsequently, the resultant product was centrifuged ($\omega$ = 8000 rpm) several times with MilliQ water, until the pH became neutral. Then, it was dried in oven at 60 °C. A final calcination step at 400 °C, under oxygen flux (6 h, 9 NL h$^{-1}$) followed to form a greyish or whitish precipitate (according to the different coverage degree of the graphene oxide surface). For the sake of comparison, pure SnO$_2$ has been prepared through the same synthetic route, without the addition of graphene oxide.

**Powders physico-chemical characterizations**

X-Ray Diffraction (XRD) analyses were performed on a Philips PW 3710 Bragg-Brentano goniometer equipped with a scintillation counter, 1° divergence slit, 0.2 mm receiving slit and 0.04° Soller slit systems. We used graphite-monochromated Cu K$\alpha$ radiation (Cu K$_{\alpha1}$ $\lambda$ = 1.54056 Å, K$_{\alpha2}$ $\lambda$ = 1.54433 Å) at 40 kV × 40 mA nominal X-rays power. Diffraction patterns were collected between 10° and 80° with a step size of 0.1°. A microcrystalline Si-powder sample was employed as a reference to correct instrumental line broadening effects.

Raman spectra were taken on a Renishaw InVia micro-Raman Spectrometer. A 50 mW, 532 nm diode laser was used for excitation. The spectrometer was equipped with a Nikon 50× objective lens (WD = 17 mm, NA = 0.45), which produced a focal spot of 1 μm$^2$ and a total power of 0.71 mW from the objective. All spectra were processed to remove cosmic rays using the inbuilt software package Wire 4.2. Raman spectra of graphite and graphene oxide samples have been deconvoluted in eight and five modes, respectively, by using the Lorentzian function and the intensity ratios between D and G bands have been calculated according to Atchudan *et al.*[37]

The BET surface area was determined by a multipoint BET method using the adsorption data in the relative pressure ($p/p_0$) range of 0.05–0.20 (Coulter SA3100 apparatus). Desorption isotherms were used to determine the total pore volume using the Barrett-Joyner-Halenda (BJH) method.

The morphology was investigated by using a Zeiss Ultraplus (filed-emission scanning electron microscopy, FESEM) at 3 kV coupled with an Energy Dispersive X-ray spectrophotometer (EDX) for the elemental analysis. Transmission Electron Microscope (TEM) analyses were performed on Hitachi H7100FA at 100 kV. The TEM grids were prepared dropping the dispersed suspension of nanoparticles in ethanol onto a holey-carbon supported copper grid and drying it in air at room-temperature overnight.

Thermogravimetric analyses were carried out by means of Metter Toledo Star and System TGA/DSC 3+ under air atmosphere (5 °C min$^{-1}$ from 30 to 800°C).

X-ray Photoemission Spectroscopy (XPS) data were collected in a Thermofisher Kratos Axis Supra photoelectron spectrometer at the Central Analytical Research Facility of the Queensland University of Technology (Brisbane, Australia). The apparatus is equipped with a monochromated Al k$_\alpha$ source (1486.7 eV), and the spectra were calibrated with respect to their Fermi level. Survey spectra were acquired at pass energy 160, high resolution spectra at pass energy 20.

To evaluate powders optical band gaps by Kubelka-Munk elaboration, Diffuse Reflectance Spectra (DRS) were measured on a UV/Vis spectrophotometer Shimadzu UV- 2600 equipped with an integrating sphere; a "total white" $BaSO_4$ was used as reference. The porosity of $SnO_2$ nanoparticle networks of the films was estimated from the optical density and SEM visible thickness as suggested by Bo et al.,[38] by adopting an absorption coefficient of $3.08 \times 10^7$ m$^{-1}$ (at 312 nm for all the powders).

**Deposition on Pt-Interdigitated Electrodes (Pt-IDEs)**

Powders were deposited on glass substrates topped with interdigitated Pt electrodes (Pt-IDEs) by a simple hot-spray method reported in our previous work.[28] Therefore, the tested IDEs were prepared by adopting pristine $SnO_2$, hybrid 4:1 and 32:1 $SnO_2$/GO powders.

**Photodetectors measurements and VOCs sensing tests**

For photodetector tests, photo- and dark-currents were measured at 25 °C with an LCS-100 Series Small Area Solar Simulator (Newport Co.). The electrode active surface was equal to 0.4 cm$^{-2}$ and the irradiation power at 312 nm was 1.5 μW cm$^{-2}$. The responsivity and detectivity were, then, calculated according to the equations reported elsewhere.[38] Regarding ethanol (EtOH), acetone and ethylbenzene (EtBz) sensing, $O_2$ (BOC Ltd) and $N_2$ (BOC Ltd) were controlled by mass flow controller (Bronkhorst), with a total gas flow rate of 0.5 L min$^{-1}$. The target gas (10 ppm in $N_2$, Coregas) were diluted to 1 ppm and lower concentrations by using the simulated air (0.1 L min$^{-1}$ $O_2$ + 0.4 L min$^{-1}$ $N_2$, BOC Ltd) before purging into the chamber, keeping constant the total flow rate. The temperature of the hotplate in the gas sensing chamber (Linkam) was controlled by a temperature controller and, when the operating temperature was lowered (equal or below 150 °C), UV light was also exploited. Therefore, the samples were illuminated through a quartz window by a solar simulator (NewSpec, LCS-100) with an FGUV5-UV – Ø25 mm UG5 Colored Glass Filter (AR Coated: 290 - 370 nm, Thorlabs Inc). For the gas sensing tests, two gold probes were separately placed on top of the powders covered IDEs, and the dynamic response was recorded by an electrochemical workstation (CHI 660E, USA), applying a bias of +1.0 V. The sensor response is reported as: $(R_{air} / R_{analyte}) - 1$, where $R_{air}$ is the film resistance in air and $R_{analyte}$ is the film resistance at a given concentration of the target gas.[10] Both sensors response and recovery times have been evaluated considering the 90% of the final response.[28]

**Results and Discussion**

**Synthesis of $SnO_2$-GO Nano-heterojunction Network**

The graphite conversion into graphene oxide material and the subsequent formation of nano-heterojunctions with a three-dimensional $SnO_2$ network have been verified by a combination of physical and chemical characterizations. Two different relative GO concentrations of $SnO_2$/GO 4:1 and 32:1 were investigated to evaluate the potential optoelectronic and chemoresistive performances of the $SnO_2$-GO nano-heterojunctions. These $SnO_2$-GO ratios were selected with respect to the previous study on ZnO-GO nanoscale heterojunctions.[29] Figure 1a–b shows a comparison of the X-Ray Diffraction patterns and Raman spectra of the pristine graphite and GO, along with the structural data of bare $SnO_2$ and the synthesized $SnO_2$-GO composites. The effective transformation of graphite material into graphene oxide was assessed by both the main GO diffraction peak (at 2θ value of 12°) and the intensity increase of the ratio between the D and G Raman bands up to 1.01 of GO vs 0.25 of the precursor graphite (Fig. 1a–b, red spectra).[28,29] Indeed, during the oxidation process oxygen functional groups are introduced into the graphitic chain, causing either an increase of the D band intensity[28,39] or a small shift (ca. 25 cm$^{-1}$ upwards for G band and 45 cm$^{-1}$ downwards for D band) of the bands positions,[28] due to the achievement of a highly defective structure.[36,40,41] Moreover, the gradual integration of GO nanodomains into the $SnO_2$ matrix was revealed by both the presence of Raman bands relative to the graphene oxide material (Fig. 2b, blue and violet spectra) and the very small XRD crystallites diameter (ca. 5-10 nm; Fig. 1a and Table 1, 4$^{th}$ column) with respect to crystal structure of $SnO_2$[42,43] (80 nm, Table 1). The crystallite size of the $SnO_2$-GO nano-heterojunctions resembles much more the graphene oxide one (11 nm, Table 1), underlining the effective integration of the carbonaceous material into the metal oxide network.

Thermogravimetric analysis reveals that the pristine $SnO_2$ and the hybrid samples are very stable (Fig. 1c) with a mass loss of only ~4% up to temperatures of 800 °C. On the contrary, pure GO (Fig. 1c, red line) decomposes in several stages, ascribable to different processes, such as i) the loss of moisture and interstitial water between 60 and 110 °C, ii) the pyrolysis of labile oxygen-containing groups with the generation of CO, $CO_2$ and water[44] at 200 °C and iii) the breakage of sp$^2$ carbon bonds at around 480 °C.[29,40] Furthermore, the presence of tin in the hybrid samples was confirmed by EDX analysis (Figure 1d). The surface composition of the $SnO_2$-GO materials was further investigated by XPS and BET-BJH analyses. The C 1s and O 1s core level high resolution spectra of the GO compound (Fig. 1e–f, red spectra) were discussed recently.[28,29,45] Besides, the C 1s region of both pure and $SnO_2$-GO compounds shows three components, referable to C–C sp$^2$ (284.75 eV), C–O/C–OH (286.20 eV) and O–C=O (289.00 eV) bonds.[46,47] Even if the last carbon peak is mostly attributed to adventitious $CO_2$, its enhanced presence in the nano-heterojunctions is indicative of GO formation.[28,29] Figure 1f shows the O 1s core level high resolution spectra, which can be deconvoluted into three components centered at around 530.75, 531.40 and 532.60 eV. These bands correspond respectively to i) lattice oxygen anions ($O^{2-}$) in the cassiterite lattice, ii) oxygen ions ($O^{2-}$ and $O^-$) in the oxygen-deficient regions, caused by oxygen vacancies, and iii) adsorbed oxygen species (especially water molecules).[48,49] The relative amount of oxygen vacancies in the $SnO_2$-GO compounds is higher than in the pure $SnO_2$, suggesting a more defective structure as a result of the graphene oxide integration into the metal oxide matrix. Furthermore, the specific surface areas of the nano-heterojunctions are close (29 and 55 m$^2$ g$^{-1}$) to that of the pure $SnO_2$ (67 m$^2$ g$^{-1}$), following a decreasing trend with the increase of the GO content ($S_{BET}$ GO equal to 30 m$^2$ g$^{-1}$; Table 1, 2$^{nd}$ column).[28] The same trend was observed for the total pores volume data (Table 1, 3$^{rd}$ column).

| Sample | $S_{BET}$ (m² g⁻¹) | $V_{tot.\ pores}$ (cm³ g⁻¹) | $<d^{XRD}>$ (nm) | $E_g$ (eV) | Film thickness (μm) | % Film porosity |
|---|---|---|---|---|---|---|
| Graphite | 11 | 0.030 | 27 | – | – | – |
| GO | 30 | 0.020 | 11 | – | – | – |
| $SnO_2$ | 67 | 0.210 | 80 | 3.6 | 1.8 ± 0.2 | 93 ± 1 |
| 4:1 $SnO_2$/GO | 29 | 0.070 | 5 | 3.0 | 3.2 ± 0.4 | 97 ± 1 |
| 32:1 $SnO_2$/GO | 55 | 0.133 | 8 | 3.4 | 1.4 ± 0.4 | 94 ± 2 |

**Table 1.** Surface area ($S_{BET}$), total pore volume ($V_{tot.\ pores}$), crystallite domain size by XRD analysis ($<d^{XRD}>$), optical band gap ($E_g$, by Kubelka-Munk extrapolation), film thickness (by cross-sectional SEM) and film porosity percentage (obtained by means of UV/Vis spectroscopy technique).

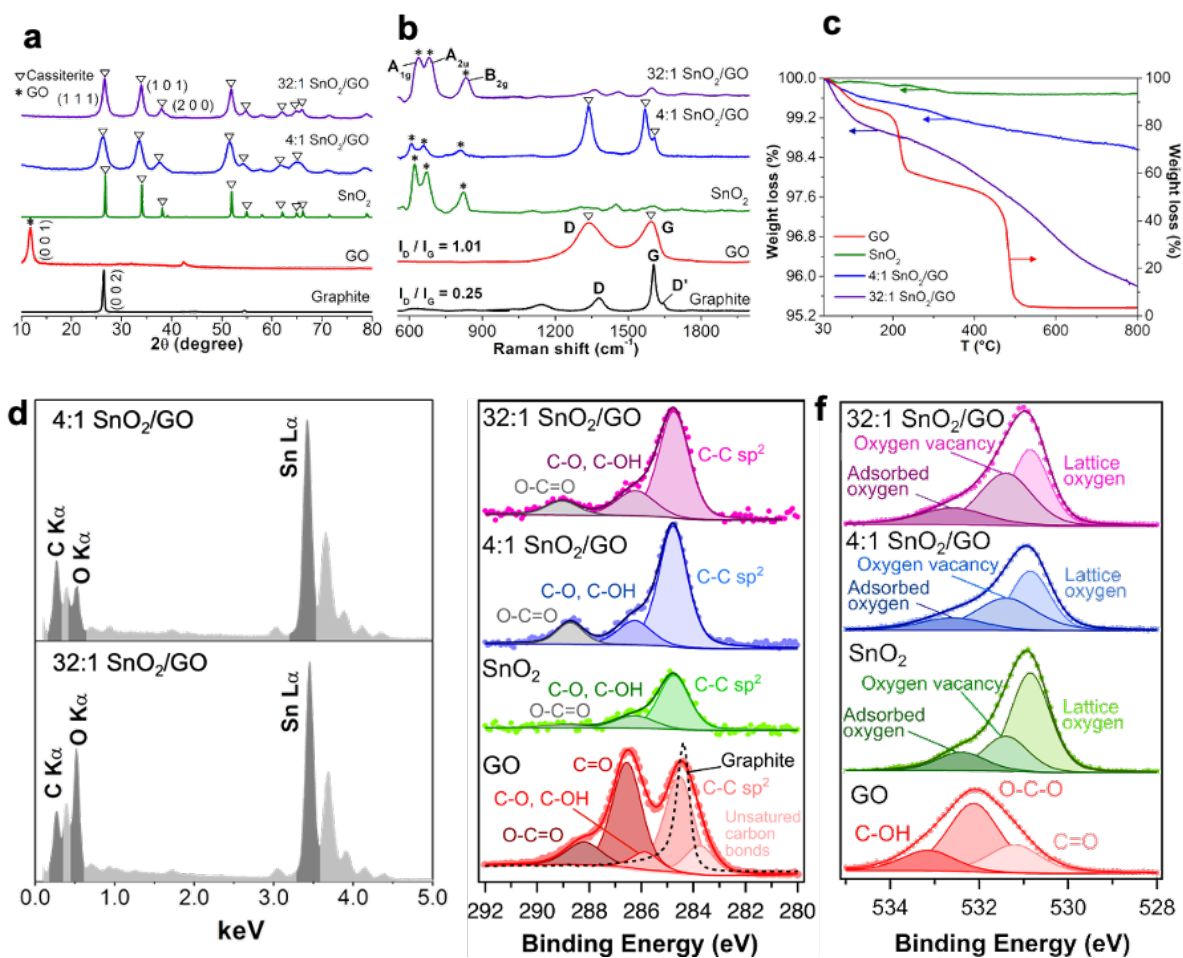

**Figure 1.** (a) XRD patterns of graphite, graphene oxide (GO), pure $SnO_2$ and hybrid $SnO_2$-GO samples (100% intensity reflection planes have been assigned to the main phases of each compound). (b) Raman spectra of all the investigated samples. (c) TGA spectra of GO, pure and hybrid nanopowders. (d) EDX spectra of 4:1 $SnO_2$/GO and 32:1 $SnO_2$/GO. XP spectra of (e) C 1s and (f) O 1s regions of graphite, GO, 4:1 and 32:1 $SnO_2$/GO.

Figure 2 shows the morphology of the pristine and composite samples by TEM and FESEM microscopy. Notably, both the 4:1 and 32:1 $SnO_2$/GO ratios seem to be comprised of spherical nanoparticles with dimensions of around 8 nm (Fig. 2b–c), which are bigger than pure oxide ones, having a size of ~5 nm (Fig. 2a).[28]

Also scanning electron micrographs display the presence of spherical agglomerates with dimensions of hundreds of nanometers for all the three $SnO_2$-based samples (Fig. 2g–i).

Overall, this set of characterization indicate that the gradual coverage of the graphene oxide sheets by $SnO_2$ is achieved, creating strong bonds between the graphene and the metal oxide nanoparticles. This tunable coverage can influence the structural and surface properties, the morphology and crystal size of the as-prepared powders, thus affecting their behavior as photo- and chemical sensing materials.

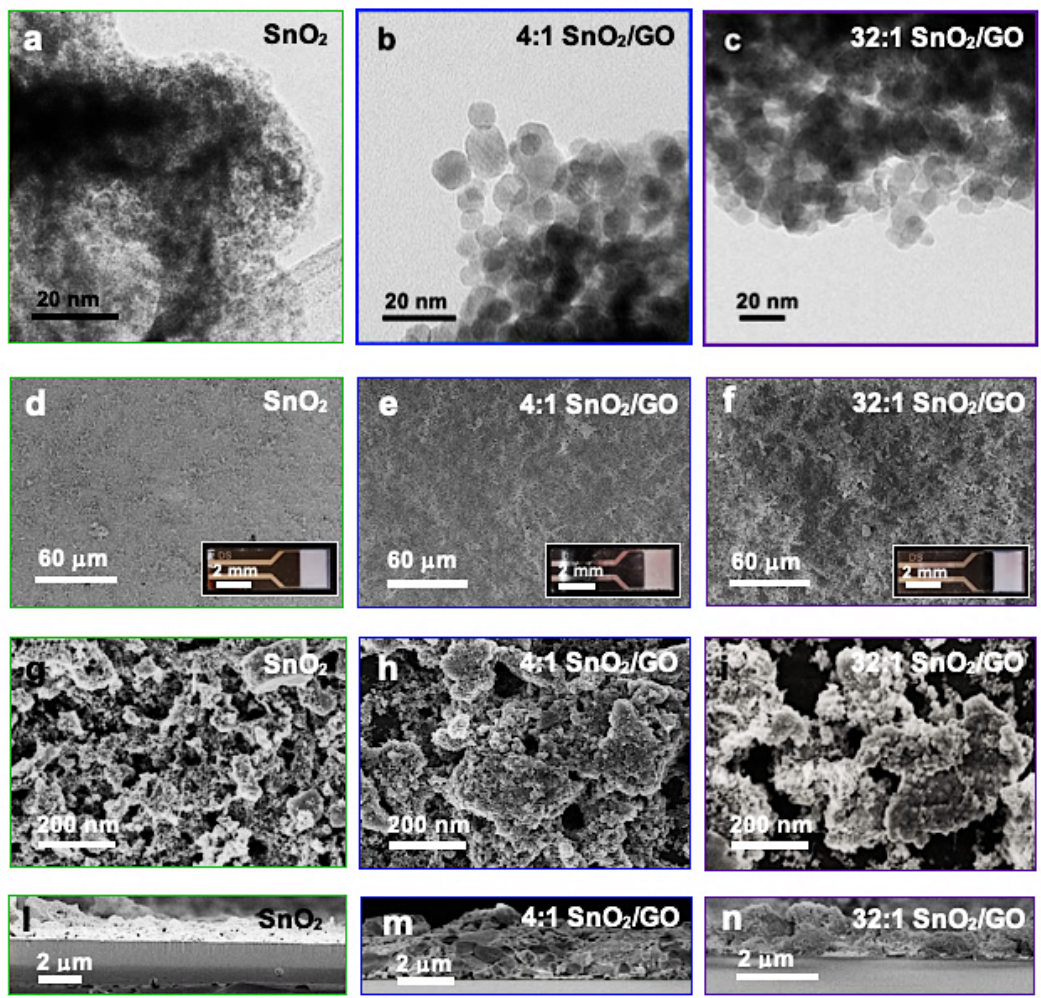

**Figure 2.** (a–c) TEM images of pristine SnO$_2$ and hybrid SnO$_2$-GO compounds. (d–i) Top view FESEM micrographs and (l–n) cross-sectional images of both pure and composite samples. Insets: photos of the relative interdigitated electrodes.

## Optoelectronic and Chemical Sensing Properties

The formation of nano-scale heterojunctions is a promising approach to improve chemical sensing at low temperatures by photo-excitation and separation of reactive electron/hole couples.[16,28,29,50] The optical properties of the tin dioxide-containing compounds were initially investigated by diffuse reflectance spectroscopy (DRS).

Figure 3a shows the Kubelka-Munk conversion of the DRS spectra, revealing similar values of 3.0 and 3.4 eV for the 4:1 and 32:1 nano-SnO$_2$/GO, respectively. These values are lower than the band gap of pure SnO$_2$ of about 3.6 eV.[28,51] This decrease is attributable to the coupling between the white tin dioxide and the brownish graphene oxide sheets. In order to investigate the powders performances, the nanopowders were deposited on Pt-IDEs via a scalable air-spraying method, obtaining highly homogeneous micrometric-thick films (Figures 2d–f). Cross-sectional FESEM images (Fig. 2l–n), reveal layers thickness of around 1.5-2.0 μm for both the SnO$_2$ and 32:1 SnO$_2$/GO nano-heterojunctions, whereas the 4:1 SnO$_2$/GO ratio resulted in a thicker film of ~3.0 μm (Table 1, 6$^{th}$ column). The estimated films porosities[29] are above 90% for all materials (Table 1, 7$^{th}$ column), in line with the values expected for aerosol self-assembly processes.[52]

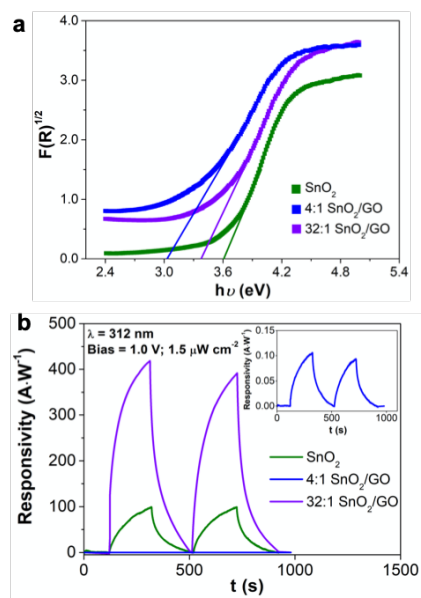

**Figure 3.** (a) Band gap values determined by Kubelka-Munk elaboration. (b) Dynamics of photodetectors responsivity for all the Sn-based compounds (λ = 312 nm, light power density = 1.5 μW cm$^{-2}$, applied bias = +1.0 V).

Insights into the optoelectronic properties and sensing mechanism of the nano-heterojunctions $SnO_2$-GO were first obtained by investigating the photo-resistive behavior under UV illumination. Lan et al.[53] proposed a high-performance UV photodetector design by combining $SnO_2$ semiconductors with three-dimensional graphene nanoflakes. The as-prepared nanocomposite films showed strong absorption in the wide UV region, owing to the presence of the 3D network that efficiently suppresses the recombination of the photo-induced electron-hole pairs and resulted in a significant enhancement of the graphene-$SnO_2$ photoresponse over that of pure $SnO_2$. Notably, the responsivity of 3D graphene-$SnO_2$ photodetector was reported to be as high as 8.6 mA $W^{-1}$ at a bias voltage of 1.0 V, which is around 8 times higher than that of the pristine tin dioxide. Here, starting from this report, the current response was acquired by applying a bias potential of 1.0 V and by UV light irradiation at 312 nm with a light power density of 1.5 μW 1<$cm^{-2}$ (Fig. 3b). The principal figures of merit for photodetectors are the magnitude of the photo-/dark-currents, responsivity and detectivity (Table 2). Especially, the last parameter quantitatively characterizes the photodetectors performances.[38] Among the investigated samples, the 32:1 $SnO_2$/GO shows the highest detectivity of 1.4 × $10^{15}$ Jones followed digressively by $SnO_2$ and 4:1 $SnO_2$/GO (Table 2, 8th column). The photocurrents, together with $I_{photo}/I_{dark}$ ratios, follow the same trend (Table 2, 3rd and 4th columns), showing a very high $I_{photo}/I_{dark}$ value of around 2400 for the 32:1 $SnO_2$/GO. Furthermore, the rise and decay times (Table 2, 5th and 6th columns) were comparable to some of the best performing $SnO_2$-based UV photodetectors.[38,54] Here, the responsivity of the 32:1 $SnO_2$/GO ratio is 400 A·$W^{-1}$ and thus very high (Table 2 and Fig. 3b) with respect to the recent literature[53]. Similarly, the 32:1 $SnO_2$/GO nano-heterojunction detectivities (Table 2, 8th column) are greater than those of some of the most performing materials.[53,55,56]

This photo-responsivity trend (32:1 > $SnO_2$ > 4:1) and the very high responsivity/detectivity, measured with the 32:1 ratio suggest a potential mechanism for the enhancement of the UV light sensing.[57] In line with the previous literature,[16,28,29,58] we suggest that a p-n type nano-heterojunction is formed between the GO, showing a p-type behavior, and n-type $SnO_2$.[33] Upon UV light illumination, photogenerated electron-hole pairs are formed, which are rapidly separated by the $SnO_2$-GO nanoscale heterojunctions, that disadvantage their recombination. This results in a higher photocurrent response especially for the 32:1 ratio, where an optimal distribution of the GO in the metal oxide nanoparticles matrix may be obtained (Figure 2). The above mechanism may also be exploited to achieve high chemical sensing at room-temperature under light-illumination. Once generated, the photoelectrons ($e^-_{hv}$) are mostly trapped on the metal oxide surface, giving rise to reactive photoinduced oxygen ions ($O_2^-{}_{hv}$). Hence, when reducing VOCs molecules are purged into the chamber, they can be oxidized by these oxygen ions releasing electrons back to the conduction band of the $SnO_2$ and thus increasing the film conductivity. To investigate the use of the optoelectronic properties of the $SnO_2$-GO nano-heterojunctions for gas sensing, here, ethanol, acetone and ethylbenzene gases were chosen as VOC model molecules. Figures 4a,d,g shows the sensor responses of the pure $SnO_2$ and the optimal 32:1 $SnO_2$/GO nano-heterojunction, as a function of both the operating temperature (OT) and the UV irradiation. Notably, at high temperature (350 °C) without UV light, both the pure and hybrid sample can detect ethanol in air down to 2 and 10 ppb concentrations, respectively. Remarkably, for an ethanol concentration of 1 ppm the signal intensity of the 32:1 $SnO_2$/GO is about three times higher than that of the $SnO_2$ (Figs. 4a and d). By decreasing the temperature to 150 and 25 °C, only the nano-heterojunction was able to sense ethanol, even if light irradiation was required (Figs. 4g and S1a), whereas no response was obtained for the bare oxide compound at room-temperature with and without light irradiation. Notably, the 32:1 $SnO_2$/GO had a very good signal-to-noise ratio down to 100 ppb at RT. Furthermore, the selectivity of the materials was investigated using different VOCs molecules. Acetone and ethylbenzene were used as alternative VOCs (Figs. 4b,e,h; 4c,f,i and S1). An analogous sensing behavior was observed for these species, achieving detection at RT of 100 ppb. However, the signals intensity was quite different (Fig. S2a), thus resulting in a possible selective detection of ethanol among the studied VOCs. Outstandingly, at RT, ethanol results in the highest signal response intensity of ca. 2 at 1 ppm, while acetone and ethylbenzene showed a lower value of about 0.3 and 0.8, respectively. This trend may follow the VOCs chemical structure, i.e. the presence of polar groups (such as hydroxyl groups) or steric hindrance (as the phenyl ring), thus leading to different affinity and reactivity with the oxide surface.[59–62] It has been previously reported that alcohols show higher sensing responses with metal oxides than aldehydes or ketones, and to a greater extent with respect to non-polar/low polar analytes, such as ethylbenzene.[61–64] Moreover, both $SnO_2$ and 32:1 $SnO_2$/GO readily respond and recover upon purging these three analytes with response and recovery times below 80 s at 350 °C (Fig. S2b and c). Reducing the operating temperature increases the response time by three/four times, depending on the VOC molecule.

A comparative summary of the $SnO_2$-GO sensing performances with literature data about $SnO_2$-based chemoresistors[4,17,18,65,66] is reported in Table 3. Interestingly, all the nano-heterojunctions, synthesized here, have superior performance than some of the best reported in the literature. In particular the 32:1 $SnO_2$/GO exhibits significantly higher signals intensity with very low limit of detection and high sensitivity, even at RT.[18,66] The feasibility of tuning the chemical response of the nano-heterojunctions by engineering their composition was further obtained correlating their sensing response at constant VOCs concentration. Figure 5 shows a comparison of responses at 1 ppm relative to different nano-heterojunctions composed of 32:1 $SnO_2$/GO, 32:1 ZnO/GO, previously reported[29], and 4:1 $SnO_2$/GO. Interestingly, the 32:1 $SnO_2$/GO has significantly higher ethanol selectivity than the other species, showing a response of about four times higher than that of acetone. These results show that, at constant relative GO amount, the tin dioxide is more selective to ethanol and has significantly higher sensitivity than the zinc oxide containing nano-heterojunction. This may be attributed to the grain boundary density of the two nano-heterojunctions. The change in material resistance depends mainly from the ratio between the grain size ($d$) and the Debye length ($\delta$).[67] If $d$ is slightly lower or equal to $2\delta$, the whole grains are depleted and change in the surface oxygen species concentration can affect the entire grain, resulting in higher sensitivity. Here, the particles sizes of both $SnO_2$-GO (~5-8 nm) and ZnO-GO (~50 $nm^{29}$) compounds are very close to twice of the Debye length of tin dioxide (~3 $nm^{66,68,69}$) and zinc oxide (~30 $nm^{70}$), respectively.

| Sample | Dark-current (nA) | Photocurrent (μA) | $I_{Photo}/I_{Dark}$ | Rise time (s) | Decay time (s) | Responsivity (A·W⁻¹) | Detectivity (Jones) |
|---|---|---|---|---|---|---|---|
| SnO$_2$ | 540 | 58 | 108 | ≈160 | ≈130 | 100 | $1.5 \times 10^{14}$ |
| 4:1 SnO$_2$/GO | 1 | 0.057 | 52 | ≈130 | ≈110 | 0.100 | $3.4 \times 10^{12}$ |
| 32:1 SnO$_2$/GO | 100 | 240 | 2380 | ≈120 | ≈100 | 395 | $1.4 \times 10^{15}$ |

**Table 2.** Figures of merit of Sn-based photodetectors (λ=312 nm, light power density, 1.5 μW cm⁻² and applied bias, +1.0 V).

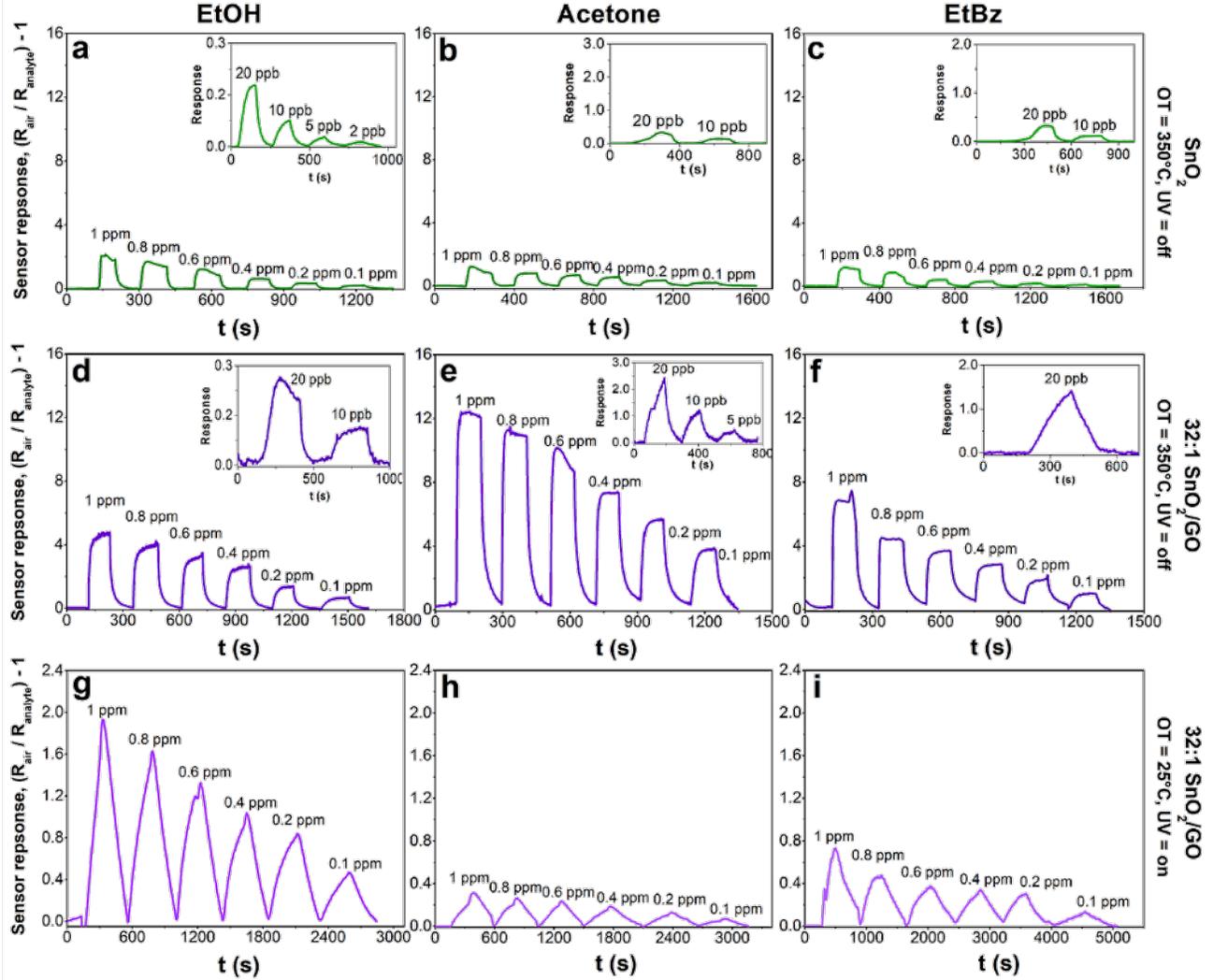

**Figure 4.** (a–c) Pure SnO$_2$ and (d–f) hybrid 32:1 SnO$_2$/GO sensors response when exposed to different low-ppm concentrations of ethanol, acetone and ethylbenzene at 350°C, without UV light. (g–i) Same tests performed with hybrid 32:1 SnO$_2$/GO materials at RT, UV-assisted. All the measurements were carried out in simulated air (20% O$_2$ – 80% N$_2$). OT = Operating Temperature.

Therefore, an improvement of the sensing behavior is expected. However, in the case of zinc oxide, Bo *et al.*[38] recently reported the further increase of ZnO nanoparticles dimensions beyond 42 nm do not help to enhance the optoelectronic features. This is mainly ascribable to the slightly greater backscattering phenomena, causing reduced photo-sensing performances. This phenomenon is reported to be typical of metal oxide semiconductors operating at low temperatures due to a greater amount of adsorbed oxygen species,[71] leading to a more hydrophilic surface. In this sample, indeed, the incomplete GO coverage results in a greater adsorption of oxygen species and moisture with respect to the 32:1 SnO$_2$/GO.

In order to demonstrate the conductivity switching at low temperature, tests at high temperatures were carried out (Fig. S3). We observed that the signals both for acetone and ethylbenzene switch from negative to positive values by increasing the OT above 150 °C, along with an increase in the relative intensities. Since this behavior was observed for ethylbenzene molecules only in the case of the 4:1 SnO$_2$/GO nano-heterojunction, it can be used as a tool to selectively sense this species at RT.

| Material | Operating temperature (°C) | VOC | Signal response, ($R_{air}/R_{analyte}$)-1[b] | LOD[a] (ppb) | Ref. | |
|---|---|---|---|---|---|---|
| Hollow SnO$_2$ | 300 | EtOH | 28.2 (100 ppm)[c] | 5000 | | |
| rGO-SnO$_2$ | 300 | EtOH | 42.0 (100 ppm)[c] | 5000 | 18 | Literature results |
| | | Acetone | 11.0 (100 ppm)[c] | – | | |
| 0.1 wt% GO/SnO$_2$ nanocomposite | 250 | EtOH | 22.5 (50 ppm) | 1000 | 17 | |
| SnO$_2$ hollow spheres | 200 | Acetone | 15.0 (50 ppm)[c] | 5000 | 19 | |
| Rh-doped SnO$_2$ nanofibers | 200 | Acetone | 59.6 (50 ppm)[c] | 1000 | 64 | |
| 3% CuO/SnO$_2$ | 280 | EtBz | 7.0 (50 ppm)[c] | 2000 of BTEX | 4 | |
| SnO$_2$ | 350 | EtOH | 2.0 | 2 | | |
| | | Acetone | 1.8 | 10 | | |
| | | EtBz | 1.5 | 10 | | |
| 32:1 SnO$_2$/GO | 350 | EtOH | 5.1 | 10 | | |
| | | Acetone | 12.5 | 5 | | |
| | | EtBz | 7.2 | 20 | | This work |
| | RT (UV) | EtOH | 2.0 | 100 | | |
| | | Acetone | 0.4 | 100 | | |
| | | EtBz | 0.8 | 100 | | |
| 4:1 SnO$_2$/GO | 350 | EtOH | 0.1 | 100 | | |
| | | Acetone | 0.6 | 100 | | |
| | | EtBz | 0.4 | 100 | | |
| | RT (UV) | EtOH[d] | 0.006 | 1000 | | |
| | | Acetone | -0.1 | 100 | | |
| | | EtBz | -0.6 | 100 | | |

**Table 3.** Comparison of SnO$_2$-based materials sensing performances towards the three investigated VOCs.

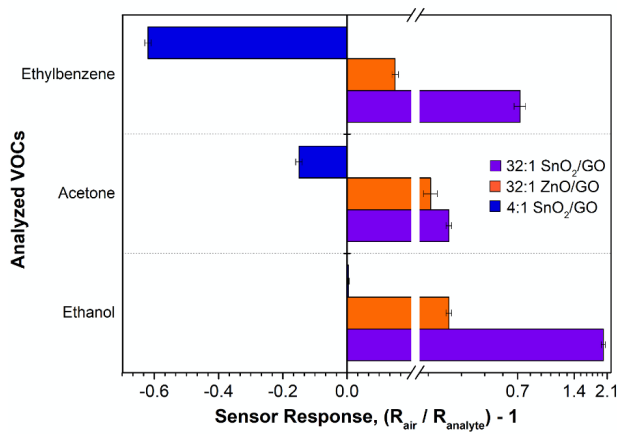

**Figure 5.** Comparison among 32:1, 4:1 SnO$_2$/GO and 32:1 ZnO/GO sensors (reported in our previous work[29] as the most perfoming sample), in terms of signal response intensity to 1 ppm of VOCs, at 25°C under UV irradiation.

As a result, tailoring of the GO content in a 3D SnO$_2$ network enables to achieve high sensitivities and selectivity towards different VOCs at room-temperature.

## Conclusions

Herein, we have presented the engineering of the optoelectronic properties of nano-heterojunctions comprising of GO domains in a three-dimensional tin dioxide matrix for the selective and sensitive measurement of VOCs at room-temperature. The nano-heterojunctions were prepared by controlling the relative graphene oxide amount in a SnO$_2$ nanoparticle network, and aerosol assembled in highly porous film structures. We observe that the tuning of the GO relative amount is a potential path to tune the selectivity to VOCs. A low GO content results in enhanced UV light responsivity of ca. 400 A·W$^{-1}$, with short 120 and 100 s rise and decay times, and room-temperature detection of down to 100 ppb of ethanol, with good selectivity against other VOCs such as acetone and ethylbenzene. A high amount of GO hinders the ethanol response at room-temperature, enhancing an opposite change of conductivity and selectivity to ethylbenzene. We propose that selectivity switching mechanism is mainly due to the different surface composition of the 4:1 SnO$_2$/GO nano-heterojunction. The latter have a highly more hydrophilic surface than that of 32:1 SnO$_2$/GO, resulting in the adsorption of moisture and hydroxyl groups at RT, and it can compete with the target VOCs for adsorption sites resulting in an opposite response to that expected for n-type semiconductors with reducing gases. The enhancement of the sensing performance over that of the bare SnO$_2$ and the tunable selectivity are attributed to the relative fraction of p(GO) – n(SnO$_2$) nanodomains, which promotes the electron-hole separation and determine the dominant surface properties. We believe that these findings can provide guidelines for the engineering of the next generation of miniaturized chemoresistive sensors for selective room-temperature detection of various VOCs. The excellent performance of the SnO$_2$-GO nano-heterojunctions as UV photodetectors provides also a tunable low-cost material for fabrication of optoelectronic devices with various applications.

## ASSOCIATED CONTENT

### Supporting Information

The Supporting Information is available free of charge on the ACS Publications website.

Figure S1: Hybrid 32:1 SnO$_2$/GO sensors response when exposed to different concentrations of (a) ethanol, (b) acetone and (c) ethylbenzene, at 150°C, with UV light (PDF).
Figure S2: (a) signal response versus different VOCs concentrations with both pure and 32:1 SnO$_2$/GO, (b) response and (c) recovery times as a function of the operating temperatures (PDF).
Figure S3: (a–c) acetone and (d–f) ethylbenzene sensing by 4:1 SnO$_2$/GO sample at 350 °C without UV light, 150 °C and room-temperature under UV irradiation (PDF).

## AUTHORS INFORMATION

### Corresponding Authors

Giuseppe Cappelletti: Dipartimento di Chimica, Università degli Studi di Milano, via Golgi 19, 20133, Milano, Italy; e-mail: giuseppe.cappelletti@unimi.it
Antonio Tricoli: Nanotechnology Research Laboratory, College of Engineering and Computer Science, The Australian National University, Canberra ACT 2601, Australia; e-mail: antonio.tricoli@anu.edu.au

### Present Address

† ARC Centre of Excellence for Future Low-Energy Electronics Technologies (FLEET), School of Physics and Astronomy, Monash University, Melbourne VIC 3800, Australia

### Author Contributions

All authors have given approval to the final version of the manuscript.

## ACKNOWLEDGMENT

The authors gratefully acknowledge Prof. Giuseppina Cerrato and Dr. Maria Carmen Valsania, Dipartimento di Chimica and NIS, Inter-department Center, Università di Torino for the acquisition of pure SnO$_2$ TEM image by means of a Jeol TEM 3010 instrument equipped with LaB$_6$ filament (operating at 300 kV). The authors gratefully acknowledge the Central Analytical Research Facility, operated by the Institute for Future Environments (QUT). Access to CARF is supported by generous funding from the Science and Engineering Faculty (QUT). A.T. gratefully acknowledges the support of the Australian Research Council DP150101939, the Australian Research Council DE160100569, the Westpac2016 Research Fellowship, the support and contribution from the ANU Grand Challenge Our Health in Our Hands. All the authors acknowledge the use of the Centre of Advanced Microscopy (CAM) at ANU.

**Table of Contents**

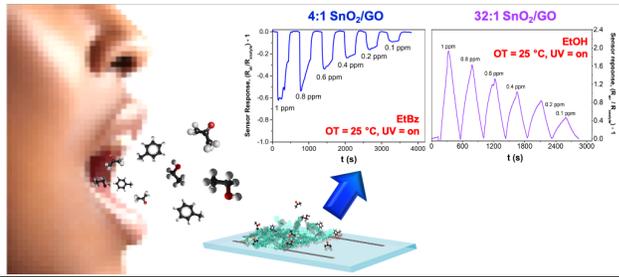